\documentclass[sigconf]{acmart}

\setcopyright{none} 

\acmYear{2017}

\usepackage{booktabs} 

\usepackage[ruled,algosection,noend,linesnumbered]{algorithm2e} 

\SetAlFnt{\small}
\SetAlCapFnt{\small}
\SetAlCapNameFnt{\small}
\SetAlCapHSkip{0pt}
\IncMargin{-\parindent}

\usepackage{etex}
\usepackage{framed}
\usepackage{multirow}
\usepackage{tabularx}
\usepackage{booktabs} 
\usepackage{adjustbox}
\usepackage{mathtools}
\DeclarePairedDelimiter{\ceil}{\lceil}{\rceil}
\DeclarePairedDelimiter\floor{\lfloor}{\rfloor}
\usepackage{amsfonts}
\usepackage{stmaryrd}
\usepackage{amsmath}
\usepackage{amssymb}
\usepackage{amsthm}
\usepackage{color}
\usepackage{float}
\usepackage[justification=centering]{caption}

\usepackage{tikz}
\usetikzlibrary{shapes, positioning, matrix, backgrounds, decorations.pathreplacing}

\newenvironment{subproof}[1][\proofname]{%
  \begin{proof}[#1]%
}{%
  \end{proof}%
}

\newtheorem{privdef}{Privacy Definition}
\newtheorem{secthm}{Security Theorem}

\newtheorem{seccjt}{Security Conjecture}

\begin{document}
\title{Mix-ORAM: Using delegated shuffles.} 

\author{Raphael R. Toledo}
\affiliation{%
  \institution{University College London}
  \country{United Kingdom}}
\author{George Danezis}
\affiliation{%
  \institution{University College London}
  \country{United Kingdom}}
\author{Isao Echizen}
\affiliation{%
  \institution{National Institute of Informatics}
  \country{Japan}}

\renewcommand\shortauthors{Toledo et al}

\begin{abstract}
Oblivious RAM (ORAM) is a key technology for providing private storage and querying on untrusted machines but is commonly seen as impractical due to the high overhead of the re-randomization, called the eviction, the client incurs. We propose in this work to securely delegate the eviction to semi-trusted third parties to enable any client to accede the ORAM technology and present four different designs inspired by mix-net technologies with reasonable periodic costs.
\end{abstract}

\thanks{This work is supported by H2020  PANORAMIX Grant (ref. 653497) and EPSRC Grant EP/M013286/1; and Toledo by Microsoft Research.}

\maketitle

\section{Introduction}
Thanks to cloud technologies, people have been able to seamlessly store impressive amounts of data on remote servers. Besides accessibility, availability and integrity, the storage providers also have to ensure to their clients the data's and the meta data's confidentiality and secure them from not only external adversaries but also from the cloud itself.
They thus employ cryptographic mechanisms to protect the communication channels, such as user authentication, data encryption and integrity checking.
These, however, do not prevent the leakage of all meta data: the servers can monitor user activities and watch which records are accessed.

Oblivious RAM (ORAM)~\cite{goldreich87}, or Oblivious Storage~\cite{boneh2011}, precisely prevents an adversary from observing the record access. In these schemes, the records are first locally encrypted and permuted in a new order before being uploaded to the untrusted cloud storage. When the user seeks a given record, the local client computes the corresponding remote index, fetches the encrypted data block and decrypts it. After a number of accesses, the database is randomized locally by the client to bring to naught any leaked information from the accesses: this is the "eviction process".

This eviction is the main bottleneck of ORAM. Indeed, the eviction consists in randomizing the whole database by permuting the records and refreshing their encryption so that an adversary loses any insight on the correspondence between the remote and local, or virtual and real, record indices. However, as we assume the number of records stored remotely to be orders of magnitude higher than what the client can store, the client has to download and randomize the database during the eviction in chunks, and do so several times so that all record ordering is equally likely. Thus as the database size grows, the eviction cost rises super linearly.

This is why we propose in this work to delegate the eviction process to dedicated semi-trusted parties. Doing so, light-weight clients could accede the ORAM technology and thanks to the use of mix networks~\cite{chaum1981untraceable} inspired designs, ORAM would become more portable.\\

In this work, we present several privacy friendly distributed systems inspired by mix-nets to safely delegate ORAM's randomization process to semi-trusted third parties.
Their advantages include the reduction of the client computation, the possibility to delay the eviction to quieter times, the guaranteed database availability during the eviction process regardless of the ORAM design and the independence from centralized parties. However careful design is required to make them scalable. Our contributions are as follows:
\vspace{.2cm}
\begin{itemize}
 \item We introduce and motivate the use of mix-net to construct delegated ORAM eviction schemes, letting very thin clients use most ORAM designs.
 \item We present a number of eviction schemes relying on mix-net, improve them with load balancing via parallel mixing.
 \item We finally evaluate the performance of our delegated eviction designs, compare them between each other and with regular eviction schemes. 
\end{itemize}

We first present the related work in Section~\ref{Related}. We then introduce the   ORAM model and how our model differ, its associated threat model and explain the different costs in Section~\ref{Prelim}. We then present two simple designs over a cascade mix-net and optimize them using random transposition shuffles over a stratified mix-net in Section~\ref{Mix-ORAM}. We then hand out our security arguments in Section~\ref{Security}, evaluate and discuss our designs in Section~\ref{Evaluation} and Section~\ref{Comparison} before concluding.

\section{Related Work}\label{Related}
\noindent\textbf{ORAM.}
ORAM was first presented by Goldreich and Ostrovsky in 1990~\cite{ostrovsky1990efficient} to prevent reverse engineering and protect software running on tamper resistant CPUs. The model was also formally extended in 2011~\cite{boneh2011} to data protection on untrusted remote clouds and in 2015 some designs were evaluated on Amazon Simple Storage Service (S3)~\cite{bindschaedler2015practicing}.
Since its introduction, ORAM enhancements have been proposed including \textit{data structures} diversification~\cite{goldreich1996software,stefanov2011towards,stefanov2013path,ren2014ring},
the use of more and more sophisticated \textit{security definitions} with statistical security~\cite{damgaard2011perfectly,ajtai2010oblivious} and differential privacy~\cite{wagh2016root}, and the revision of \textit{item lookups} with cuckoo hashing~\cite{pinkas2010oblivious} and bloom filters~\cite{williams2008building}.
Most ORAM constructions are based on a single client-server model, but multi-user designs were gradually introduced \cite{backesanonymous,franz2011oblivious,goodrich2012privacy}.\\

\noindent\textbf{Shuffling and Sorting.}
Shuffle and sorting algorithms are a thoroughly researched subject central to ORAM for the randomization process. However most of the existing methods are not useful for ORAM as they are not oblivious in that the permutations done depends on the data itself.
Examples of oblivious sorting algorithms include sorting networks such as Batcher's~\cite{batcher1968sorting} and the ones based on AKS~\cite{ajtai19830}, which unfortunately were proven to be impractical because of the high number of I/Os, but also more recent and efficient ones~\cite{paterson1990improved}.
Newer designs include the randomized Shellsort~\cite{goodrich2010randomized}, an elegant simple data-oblivious version of the Shellsort algorithm, the Zig Zag sort~\cite{goodrich2014zig} presented in 2014, the Melbourne shuffle~\cite{ohrimenko2014melbourne} and work of particular interest written by Goodrich in 2012~\cite{goodrich2012anonymous} assess the information leakage due the use of a partially compromised parallel mix-net.\\

\noindent\textbf{Mix-nets.}
Mix-nets were first presented for anonymous e-mailing by David Chaum in 1981~\cite{chaum1981untraceable}. As they became popular many improvements were made over the years~\cite{moller2003mixmaster,danezis2003mixminion,danezis2004minx,danezis2009sphinx}. Mix-nets' main goal is to give users anonymity by hiding the correspondence between the incoming users' packets and the mix-nets output. To do so, the users' messages go through several mixes which permute them and refresh their encryption. Either re-encryption~\cite{wikstrom2006adaptively} and onion encryption can be used, proofs of shuffle~\cite{groth2007verifiable,groth2007non,bayer2012efficient} and Randomized Partial Checking~\cite{jakobsson2002making} can help verify the shuffle correctness.

This work is inspired by the mix-net technology for its encryption and permutation functionalities, however, only the packet unlinkability property is of interest for ORAM. From now on, we refer traditional ORAM solutions as ORAM and our designs as Mix-ORAM.

\section{Preliminaries}\label{Prelim}
\subsection{ORAM introduction}\label{ORAM}
The Oblivious RAM system is a distributed system composed of two parties, the ORAM \emph{server} and the \emph{client}. The \emph{server} handles two data arrays, a first one we call the \emph{database} which comprises the user's encrypted records and a temporary one that we call the \emph{cache} which is of lesser size and used to hide the number of times a record was accessed. Only read and write operations are allowed on these arrays.
The \emph{client} comprises a small memory in which the cache and some additional records can fit and provides two main methods, the ORAM records access and eviction. \\

As stated in Bindschaedler's work~\cite{bindschaedler2015practicing}, most ORAM algorithms can be classified in four distinct families, the layered ORAM, the partition-based ORAM, the large-message ORAM and the tree-based ORAM, depending on the database structure and the eviction method employed. In this work we consider all algorithms that rely on periodic data-oblivious shuffle of the database. This excludes from the scope of our study solutions relying on the tree-based architecture or using higher client memory and the use of recursive algorithms such as~\cite{stefanov2011towards,shi2011oblivious} or as Path ORAM~\cite{stefanov2013path} which uses $O \left ( \log n \right )$ private memory and $O\left ( (\log n)^2\right )$ access overhead. We can now thus describe the client methods as follows.\\

\noindent\textbf{The access method:} In order to perform a read or a write operation, the client first downloads if needs be the cache and checks locally whether the desired record is present. If so, a dummy record is fetched from the database, else the client fetches the desired record. The cache is then updated with a newly encrypted version of the fetched element before finally being uploaded back to the remote ORAM server.\\

\noindent\textbf{The eviction method:} When the cache is full, the client starts the eviction process to prevent too important information leakage. The eviction consists of two parts where the client first \textit{rebuilds} the database before starting the \textit{oblivious shuffle}.

During the \emph{rebuild phase}, the client obliviously uploads back the records from the cache back to the ORAM database. After doing so, the client can finally starts \emph{the oblivious shuffle} during which chunks of the database are permuted and encrypted obliviously before being sent back to the database.\\\

Current ORAM solutions have so far relied on the client locally encrypting and shuffling the records in an oblivious manner. Batcher's sorting network~\cite{batcher1968sorting} for instance requires $O \left (n (\log n)^2 \right )$ I/Os, AKS~\cite{ajtai19830} or Zig-zag sorting networks~\cite{goodrich2014zig} which use $O\left ( n \log n \right )$ I/Os but with large constant factors or finally the Melbourne Shuffle~\cite{ohrimenko2014melbourne} which is not not based on a data-oblivious sorting algorithm, using only  $O \left ( \sqrt n \right )$ I/Os but with a large and  fixed message size of $O \left ( \sqrt n \right )$.
We propose in this paper a new oblivious shuffle performed by semi-trusted third parties, the difficulty of which being that the records must be shuffled in a scalable way without leaking information about the correspondence between indices to any party.\\
 
In this paper, we will reuse the previously defined system with the addition of the mix-net, a group of independent servers capable of encryption and permutation, and the following modification of the oblivious shuffle. 
When starting the delegated eviction, the client first selects a set of mixes which will randomize the database. It then generates and sends to them randomization instructions. The mixes use these to compute the encryption keys and permutation seeds and fetch their allocated records from the database. They then randomize the records by encrypting and shuffling them with the keys and seeds, and forward them to the next mix(es) in what we call a round. This randomization process is then done a number of times as specified in the instructions before the records are uploaded back to the database ready to be accessed.

\subsection{Security definitions and Threat model}\label{Threat}

We presume here the existence of a motivated adversary trying to subvert a target user's privacy by learning the correspondence between the remote and the local record indices. We furthermore assume that the user protects its data with an ORAM system compliant with the Privacy Definition~\ref{def:Oram} introduced by Stefanov et al.~\cite{stefanov2011towards} (see below) and additionally that all communications between the client, ORAM server and mixes are secured but may be intercepted as in the \textit{global passive adversary} assumption.
Finally, we suppose the adversary has corrupted the ORAM server and all but one mixes, and that the compromised machines behave in a \textit{honest but curious} way in that all operation are correctly performed but passively recorded and all secrets shared with the adversary.

\begin{privdef}\label{def:Oram}
Let's denote a sequence of $k$ queries by $\text{seq}_k=\{(\text{op}_1, \text{ad}_1, \text{data}_1), \text{ ... },(\text{op}_k, \text{ad}_k, \text{data}_k)\}$, where $op$ denotes a read or write operation, $ad$ the address where to process the operation and $data$ the block to write if needs be else $\perp$.
We denote by $ORAM(seq_k)$ the resulting randomized data access from the ORAM process with input $seq_k$.
The ORAM guarantees that $ORAM(seq_k)$ and $ORAM(seq'_{k'})$ are computationally indistinguishable if $k=k'$.
\end{privdef}

This work focuses on the ORAM eviction process and more precisely on the oblivious shuffle problem where sequences of data-blocks are shuffled and encrypted in order to hide the correspondance between records indices after a number of accesses has been performed. This problem refers to the eviction of the shelter in the database in the Square Root solution~\cite{ostrovsky1990efficient} and to the eviction of upper partitions in a lower ones in the Hierarchical case~\cite{goldreich1996software}. We consider the threat of a probabilistic polynomial-time (PPT) adversary and evaluate the security of our designs by looking at the information leakage of the oblivious shuffle and at the correctness of the cryptography methods used by the designs.\\
\subsection{Cryptographic Primitives}

\noindent\textbf{PRG \& Seeds.}
ORAM systems use pseudo random generators (PRG) and seeds to link remote and real indices. A distribution $\mathcal{D}$ over strings of length $l$ is said pseudo random if $\mathcal{D}$ is indistinguishable from the uniform distribution over strings of length $l$~\cite{katz2014introduction}. That means it is infeasible for any probabilistic polynomial-time adversary to tell whether the string was sampled accordingly to $D$ or was chosen uniformly at random. A PRG is a deterministic algorithm that receives as an input a short random key and stretches it into a long pseudo random stream.\\\

\noindent\textbf{Encryption.}
ORAM designs heavily rely on encryption to obfuscate the records during the eviction and the  access. In this work we will use symmetric encryption for its  rapidity and also public encryption for the key and seeds derivation.
The Advanced Encryption Standard (AES)~\cite{daemen2013design} has high speed and low RAM requirements: it has throughput over 700 MB/s per thread on recent CPUs such as the Intel Core i3~\cite{mcwilliams2014hardware} which makes it the ideal choice for ORAM.
We also make use of elements of a elliptic curve group of prime order satisfying the decisional Diffie-Hellman assumption to compress the instructions sent to the mixes.

\subsection{Model}\label{model}

\noindent\textbf{System.} We consider in this work an ORAM remote server consisting of a database with memory of $n\ b$-bit long data blocks and a cache with memory of $s,\ s\ll n,\ b$-bit long data blocks. We furthermore consider a mix-net composed of $m$ mixes, and a client with memory of $s$ data blocks. The ORAM server, the mixes and the client additionally have a small memory of capacity $\mathcal{O}(m)$ to store extra information about the permutation and encryption. We consider facing the threat of a PPT adversary and call $\kappa$ our security parameter representing the length of our encryption keys and permutation seeds that we denote by $k$ and $\sigma$ respectively.\\

\noindent\textbf{Costs.} We are interested on one side in the costs incurred by the client for recovering a record index, for decrypting a record and the extra space needed, on the other side in the total costs incurred by the mixes encrypting the records, permuting them and the transferring them.
Some operations can be preprocessed by the mixes while the records are being transferred, as the key and seeds generation and the record allocation, and as thus will not be the main focus.

\section{Mix-ORAM}\label{Mix-ORAM}
This work aims at obviously sorting the database from an old state $\Pi_{\sigma}(DB)$ to a new one $\Pi_{\sigma'}(DB)$ with the aid of a mix-net. As the seed space is not structured, it is a NP-hard problem to find for any $\sigma_1$ another seed $\sigma_2$ such that $\Pi_{\sigma_2} \circ \Pi_{\sigma_1} = I$, the overall mix-net is limited to perform a permutation $\Pi_{\sigma''}$ such that $\Pi_{\sigma'}(DB) = \Pi_{\sigma''} \circ \Pi_{\sigma} (DB)$. 
We present two ways to do the oblivious shuffle. We call the first way the \emph{Layered method} which consists in having the mixes permute the records with independent random seeds, i.e. the permutation layers are simply stacked, and the client storing the indices. We call the second way the \emph{Rebuild method} where the mixes obliviously undo the permutations done at the previous done, $\Pi_{\sigma}$, before shuffling the records with new random seeds. 

In this section, we first introduce the two methods to randomize the records during the eviction over a simple cascade mix-net. We then optimize the Mix-ORAM schemes by considering a stratified mix-net together with distributed shuffle algorithms.

\subsection{A simple Mix-ORAM}\label{ASMO}
\begin{figure*}
\begin{tikzpicture}[auto, semithick, node distance= 4em]
\tikzstyle{every state}=[fill=white,draw=black,thick,text=black]
\node[draw, cylinder, shape border rotate=90, minimum height=3em,minimum width=2em,  label={[align=center, yshift=0.5em]\textbf{Status} \\$k_0, (\sigma, k_1)$}]    	(X)				  {DB};
\node[draw, rectangle, align=center, label=below:{\color{white}'\color{black}$k_{1,1}$\color{white}'}, label=above:{${\sigma_{1}}$}]   		(A)[right of=X, yshift=1em]   {$M_1$};
\node[draw=none, fill=none] (XX)[right of=A, xshift=-1.25em] {...};
\node[draw, rectangle, align=center, label=below:{\color{white}'\color{black}$k_{1,m}$\color{white}'}, label=above:{${\sigma_{m}}$}]   		(AA)[right of=XX, xshift=-1.25em]   {$M_m$};
\node[draw, rectangle, align=center, label=below:{$k_{0,1}',\ k_{0,m}$}]    		(B)[right of=AA, xshift=.25em]    { $M_m$};
\node[draw=none, fill=none] (XXX)[right of=B, xshift=-1.25em] {...};
\node[draw, rectangle, align=center, label=below:{$k_{0,m}',\  k_{0,m} $}]    		(BB)[right of=XXX, xshift=-1.25em]   { $M_1 $};
\node[draw, rectangle, align=center, label=below:{$k_{1,m}'$}, label=above:{$\sigma_{m}'$}]    		(C)[right of=BB, xshift=.25em]   {$M_m$};
\node[draw=none, fill=none] (YY)[right of=C, xshift=-1.25em] {...};
\node[draw, rectangle, align=center, label=below:{$k_{1,1}'$}, label=above:{$\sigma_{i=1}'$}]    		(CC)[right of=YY, xshift=-1.25em]   {$M_1$};
\node[draw, cylinder, shape border rotate=90, minimum height=3em,minimum width=2em, label={[align=center, yshift=0.5em] \textbf{Status}\\$k_0', (\sigma', k_1')$}]    	(Y)[right of=CC,yshift=-1em]	  {DB};

\draw[decoration={brace, amplitude=1em}, decorate]  ([yshift=1em, xshift=-.75em]A.north west) -- ([yshift=1em,xshift=.75em]AA.north east) node[above, pos=0.5, yshift=1em, align=center] {(1) Unwrapping \\ D/$\Pi^{-1}$ phase};
\draw[decoration={brace, amplitude=1em}, decorate]  ([yshift=1em, xshift=-.75em]B.north west) -- ([yshift=1em,xshift=.75em]BB.north east) node[above, pos=0.5, yshift=1em] {(2) Simple E/D phase};
\draw[decoration={brace, amplitude=1em}, decorate]  ([yshift=1em, xshift=-.75em]C.north west) -- ([yshift=1em,xshift=.75em]CC.north east) node[above, pos=0.5, yshift=1em, align=center] {(3) Wrapping \\ E/$\Pi$ phase};

\path[->]
([yshift=1em]X.east) edge     	node{}    	(A.west)
(A.east) edge     	node{}    	(XX.west)
(XX.east) edge		node{}		(AA.west)
(AA.east) edge		node{}		(B.west)
(B.east) edge		node{}		(XXX.west)
(XXX.east) edge		node{}		(BB.west)
(BB.east) edge     	node{}    	(C.west)
(C.east) edge     	node{}    	(YY.west)
(YY.east) edge     	node{}    	(CC.west)
(CC.east) edge     	node{}    	([yshift=1em]Y.west);
\end{tikzpicture}
\centering
\caption{Cascade Mix-ORAM.\newline Rebuild method (all phases) and Layered method (only the Wrapping phase) } \label{fig:eviction}
\end{figure*}
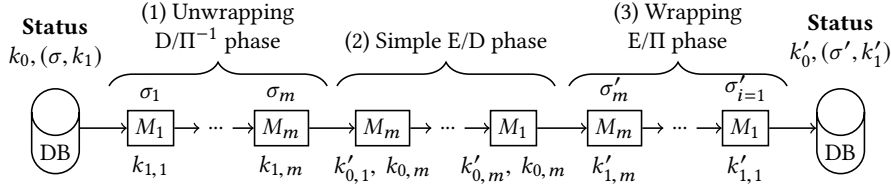

We introduce here the two randomization methods over a semi-trusted cascade mix-net, a topology in which the mixes receive and process a batch of packets sequentially has shown in Figure~\ref{fig:eviction}.
For each method, we show how the mix-net encrypts and permutes the records and how the client recovers a record plain text.

\subsubsection{Cascade Layered scheme.}\label{section:CL}
In this scheme, we use the the layered encryption method over the cascade mix-net for the eviction of the database. The underlying principle of the layered method is to have the whole database go through the mix-net once, with each mix independently encrypting and permuting the records. This method corresponds to the sole Wrapping phase (3) of Figure~\ref{fig:eviction}.

We assume in the Layered method that the records were preprocessed before being uploaded to the ORAM server as follows. Each record is first appended with its current index used as label and an initialization (IV) token as shown in Figure~\ref{fig:ldata}. The resulting data structure is then encrypted in two stages with AES in CBC mode. The label and record are first encrypted together using the IV token as initialization vector then the IV token is encrypted with the first bits of the label-record cipher. All the data structures are then permuted and finally uploaded to the ORAM server, while their indices are locally stored on the client.

\vspace{0.2cm}

\begin{table}[H]
\begin{minipage}[t][][b]{.475\textwidth}
\vspace{0pt}
\centering
\begin{tabular}{|c|c|}
IV token  &  label || record\\
$8\cdot\ceil{\log(n)/8}$ bit  &  $8\cdot\ceil{\log(n)/8}$ +$b$ bit   \\
\end{tabular}
\centering
\caption{Layered method data structure.}
\label{fig:ldata}
\end{minipage}
\end{table}
 
In the following, we present the \emph{Mix instructions} sent to the mix-net, used for retrieving the records from the ORAM database and compute the permutation seeds and encryption keys and the \emph{Mix operations}. The client decryption and access methods are then detailed in the \emph{Client operations}.\\

\noindent\textit{Mix instructions.}
To start the eviction, the client sends to each mix $M_i$ the ordered list of mixes $list = (ports,\ ips)$ involved in the oblivious shuffle, the database access information $db$, the security parameter $\kappa$, and $\alpha_i$, an element of a cyclic group of prime order satisfying the decisional Diffie-Hellman Assumption.
$$ C \rightarrow M_i\ :\ db,\ \alpha_{i},\ \kappa,\ list $$

Let $g$ be a generator of the prime-order cyclic group $\mathcal{G}$ satisfying the Diffie-Hellman Assumption and $q$ the prime order of $\mathcal{G}$. We assume that each mix $M_i$ has a public key $y_i=g^{x_i}\in \mathcal{G}^*$ with $x_i \in_{\mathbb{R}} \mathbb{Z_q}$ being its private key. We also assume that the list of $(mix_i, y_i)$ is distributed in a authenticated way thanks to a Public Key Infrastructure (PKI).
To generate the $\alpha$s, the client pick at random in $\mathbb{Z}_q$ for each mix $M_i$ the element $z_i$. The group elements and mixes' private keys are used to generate the shared secrets $ss$ from which the encryption keys and permutation seeds are derived with the aid of the HKDF derivation function~\cite {krawczyk2010cryptographic} as follows:
\begin{align*}
\alpha_i &= g^{z_i}\ ;\ ss_i = y_i^{z_i} = \alpha_i^{x_i} \ ;\ k_i,\ \sigma_i=\text{hkdf}(ss_i,\ \kappa)
\end{align*}

\noindent\textit{Mix operations.} The mix $M_i$ first decrypts the Mix instructions, generates the encryption key $k_i$ and the permutation seed $\sigma_i$. The mix then receives the records from the mix $M_{i-1}$ or fetch them if it is the first mix in the list. It then encrypts all the data structures with the new encryption key $k_i'$ as said previously and permutes them with the new seed $\sigma_{i}'$ before sending it to $M_{i+1}$ or the database if $M_i$ is the last mix in the list.\\

\noindent\textit{Client Operations.} The client can find the record index locally as it was stored previously. To decrypt a record, the client uses a trial and error recursive algorithm : the client first decrypts the data-block successively with all the shared secrets set by the latest Mix instructions and decrypts it another time with its private key. If the label is the record index, the process stops and the record is returned. If not, the data-block is re-encrypted with the client's private key and the algorithm restarts with the newly encrypted record and the shared secrets used in older Mix instructions.

We moreover modify the \emph{Access method} to prevent timing attacks as follows. When accessing a record, the client now directly encrypts the data-block with its own private key and updates it in the local cache. The client then uploads the cache to the remote server.
After doing so, the client can perform the read/write operation: it either overwrites the record with its new version ; either decrypts the record with its private key and then starts the trial and error algorithm. Once the plain text is retrieved, the client finally encrypts the record a last time with its private key and stores it locally and at the next eviction overwrites the cached version with it. \\

\noindent\textit{Costs.}
As the whole database is sent through the mix-net, the mix communication cost is $ (m+1) \cdot n \cdot b$, the mix permutation cost is $mn C_{\Pi}(n)$ with $C_{\Pi}(n)$ the cost of permuting $n$ elements and the encryption cost $m n C_{cbc}$ with $C_{cbc}$ the cost of encrypting one data block. The client Lookup cost is of the order $O(1)$ thanks to the $n$ indices stored locally for a total of $n\log(n)$ bits. We will discuss of the decryption cost in the Evaluation Section~\ref{Evaluation} as it depends on the average number of encryption layers, however $2\kappa m$ bits are used to store locally the group elements given that we always blind the same $m$ elements.

\subsubsection{Cascade Rebuild scheme.} \label{section:CR}

The rebuild method aims at replacing all the mix encryption and permutation layers with new ones ; the key challenge here is that the intermediaries should  never see the underlying client records. 
In order to achieve this, the records are encrypted and decrypted in two phases : a simple encryption-decryption ((2) E/D phase) and then an encryption-permutation ((1) the Unwrapping phase and (3) the Wrapping phase) as shown in Figure~\ref{fig:eviction}. We use in the Rebuild method the AES encryption method in Counter mode and take as counter the record current index.

Before uploading the records to the untrusted storage for the first time, the client prepares the data as follows. The records are first encrypted with the client own private keys. The first encryption keys and permutation seeds are then generated and used to encrypt the records once with fixed counters and another time while permuting the records at the same time (with varying counters), i.e. locally doing the Simple Encryption phase (2) and the Wrapping phase (3) of Figure~\ref{fig:eviction}.\\

\noindent\textit{Mix instructions.}
The client sends to each mix the same information as in the Cascade Layered design but with two group elements : $\alpha_{i}$ being used to undo the old permutations and decrypt the old encryption layers (in the Unwrapping and E/D phase), and $\alpha_{i}'$ used for the new encryption and permutations (in the E/D phase and the Wrapping phase). The client thus send to each mix $M_i$:
$$C \rightarrow M_i\ :\ db,\ \alpha_{i},\ \alpha_{i}',\ \kappa,\ list $$

The mix $M_i$ thus computes the permutation seeds and encryption keys as follows.
\begin{align*}
\alpha_i &= g^{z_i},\ ss_i = y_i^{z_i},\ k_i, \sigma_i=hkdf(ss_i,\ \kappa) \\
\alpha_i' &= g^{z_i},\ ss_i' = y_i^{z_i'},\ k_i', \sigma_i'=hkdf(ss_i',\ \kappa)
\end{align*}

\noindent\textit{Mix operations.} In this scheme, the mix $M_i$ receives a list of encrypted records from the mix $M_j$ or fetch the database if it is the first mix. During the Unwrapping phase, the mixes remove first the old encryption and then the permutations thanks to the old keys $k_{i}$ and seeds $\sigma_{i}$  and send the records to $M_{i+1}$, the last mix $M_m$ instead sends them to itself. The mixes then in the simple E/D phase encrypt the records with both the new and old keys thanks to AES in Counter mode commutativity and send to the previous mix in the list, with the first mix $M_0$ sending it to the last mix $M_m$. Finally, in the Wrapping phase, the records are permuted with $\sigma_{i}'$, and then encrypted with $k_{i}'$ and sent to $M_{i-1}$ or the database for the first mix.\\

\noindent\textit{Client operations.} To find a record index, the client uses the last $m$ seeds to simulate the mix permutations during the previous eviction or the preprocess. 

When retrieving a record, the client first computes the record's remote index using the permutation seeds. The client saves all intermediary and final indices and use them as counters to decrypt the record sequentially $r$ times. The client then decrypts the record with all the shared secrets and its own encryption key together with the original index as counter to reveal the plain-text. The client then updates the encryption of the record to read or write in the local cache, and uploads the cache back to the ORAM server.\\

\noindent\textit{Costs.}
As the whole database is sent through the mix-net three times, the mix communication cost is $ (3m) \cdot n \cdot b$, the mix permutation cost is $2mn C_{\Pi}(n)$ with $C_{\Pi}(n)$ the cost of permuting $n$ elements and the encryption cost $4 m n C_{ctr}$ with $C_{ctr}$ the cost of encrypting one data block. The client Lookup cost is of the order $m C_{\Pi}(n)$. The client decryption cost is $2mC_{ctr}$, and the group elements stored on the client requires $2\kappa m$ bits of storage.\\

Both of the Cascade Layered and Cascade Rebuild designs are not efficient as they do no fully utilize the mixes' capacity: for a single user only one mix works at a time. However, the designs can be used in pipeline when dealing with several users.

To increase the mix-net efficiency, we next study in the following section parallelization to distribute the workload among mixes while keeping the shuffle oblivious. To do so, we change the mix-net configuration to a stratified one and introduce \emph{random transposition shuffles}.

\subsection{Parallelizing the Eviction process.}\label{Parallel}
From here on, we replace the cascade configuration of the mix-net with a stratified one and have the mixes simulate random transposition shuffles (RTS) thanks to the use of private and public permutations. We also calculate the number of rounds needed to reach good security by presenting firstly the mixing time of $k$-RTS before introducing ORAM assumptions to reduce the expected time to achieve randomness.

\subsubsection{$k$-Random Transposition Shuffle.}\label{kRTS}
Random Transposition Shuffles (RTS) are widely used models in the study of card shuffling. It consists in a player picking randomly a couple of cards from a same deck, permuting them according to a coin toss and putting them back at the same location.
These steps, usually called a round, are then repeated until the deck of cards has been properly shuffled, i.e. until every card sequence is equally possible.

RTS are natural candidates for amortized ORAMs : the rounds are independent and can be run by different entities over time. 
Diaconis et al. in 1986~\cite{aldous1986shuffling} have proved that the RTS mixing time of a deck of $n$ cards is of the order  $O\left(n\log n \right)$, we first look at oblivious $k$-RTS, an RTS where the client picks and transposes locally $k$ distinct cards to make the scheme more efficient. We stress the difference between doing successively $k/2$ transpositions and what we call $k$-RTS: in the first case, an element can be transposed several times in a row of $k/2$ transpositions while in $k$-RTS it is transposed at most once. The result we present affirms that  $k$-RTS converges to the uniform distribution more rapidly than repeating normal RTS.  

\begin{secthm}
\textbf{Mixing time of $k$-RTS.} A $k$-random permutation shuffle of a $n$ card game reaches the uniform distribution in $\tau$ rounds, such that
$$E(\tau) < \frac{2 n}{k}\cdot \log(n)$$
\begin{proof}
See Appendix~\ref{proof:kRTS}.
\end{proof}
\end{secthm}

\textbf{Remark}. This theorem gives an upper bound of the number of rounds for $k/2$ disjoint transpositions. However, we use in practice PRG keys which do not guarantee that $k/2$ transpositions are done. The permutation done with the PRG can be decomposed as a sequence of transpositions which may not be disjoint or of size $k/2$. We nevertheless consider that in practice an oblivious $k$-RTS implies computation and communication cost of the order of $\mathcal{O} \left(\frac{n}{k}\cdot \log(n)\right)$.\\

To simulate the $k$-RTS over the stratified mix-net we will allocate to each mix a range of indices, for instance the mix $M_i$ fetches from the database the records whose indices are comprised in $\llbracket i\cdot n/m : (i+1)n/m -1 \rrbracket$. Each mix then fetches its allocated records and permutes them locally. Finally, all mixes perform the same public permutation on all the indices to allocate the records in the next shuffling round and forward the records to the mixes accordingly. This last permutation is required to simulate the random card choice of the classic RTS shuffle.

\begin{algorithm}
\SetKwComment{Comment}{//}{}
\DontPrintSemicolon
\KwIn{Public seeds $\sigma_{pub, rnd}$;\\ \qquad Number of records $n$;\\\ \qquad Number of mixes $m$;}
$records \gets {\Pi}_{\sigma_{pub, rnd}}(\llbracket 1 : n \rrbracket)$;\\
$alloc \gets []$;\\
\ForAll {$i \in \llbracket 1, m \rrbracket$}
{
	$alloc \gets alloc \cup records[i\cdot n/m : (i+1)\cdot n/m]$;\\
	$alloc[i]\gets [alloc[i][j]$ for $j \in [1:n/m]$ if $alloc[i][j] \in [idx\cdot n/m : (idx+1)\cdot n/m] ]$;\\
}
\KwOut{$alloc$}
\caption{Public Record Allocation for mix $M_{idx}$ at round $rnd$}
\label{alg:PRA}
\end{algorithm}

When $m$ mixes perform in parallel the $k$-RTS, we can improve in theory by another factor $m$ the eviction computation time. However to guarantee that no information is leaked to the adversary, we need each honest mix to perform $r=2 m\log n$ rounds, hence we ask each mix to perform the $k$-RTS for $r$ rounds.
\subsubsection{Oblivious Merge}\label{OM}
Before the eviction algorithm is run, the database can be divided in two sets of records depending on whether or not they were retrieved by the user. As such, the database can be represented as a simple binary array of $n$ bits out of which $s$ are 1s, the accessed ones, and $n-s$ are 0s, the others.
We argue that in this representation, elements of the same sets are indistinguishable to the adversary thanks to prior encryptions and permutations and thus, fewer rounds are necessary to obliviously shuffle the database from this state since we only need to hide from which set the records are from. Indeed, this assumption significantly reduces the number of possible orders in the adversarial view, there are ${n \choose s}$ orders instead of $n!$ (using the Stars and Bars theorem~\cite{feller1950probability}).

We now consider the RTS process in that scenario and assume the records (the bits) are re-encrypted before being permuted such that the merge of the two sets is oblivious to the adversary.

\begin{secthm}
An oblivious merge (OM) of 2 indistinguishable sets of respective size $n$ and $s$ elements requires $\tau$ rounds of 2-RTS such that any arranging is possible, with
$$\tau(\epsilon) \leq \frac{n}{2}  \cdot \log \left (\frac{n}{s}\right)$$
\begin{proof}
See Appendix~\ref{proof:OM}.
\end{proof}
\end{secthm}

The $k$-RTS decreased the mixing time by at least a factor $k$, and does so independently of the items to shuffle, we make the following conjecture.

\begin{seccjt}\label{sec:kOM}
A $k$-oblivious Merge ($k$-OM) of 2 indistinguishable sets of $n$ and $s$ element requires $\tau$ rounds such that any order is equally possible, with
$$ \tau(\epsilon) \leq \frac{n}{2k}  \cdot \log \left (\frac{n}{s}\right) $$
\end{seccjt}
\subsection{Parallel Mix-ORAM}\label{parallelMixORAM}
\begin{figure*}
\centering
\begin{tikzpicture}[auto,
thin, 
scale=0.5,
block/.style={draw, fill=white, rectangle, font=\small}]

\node[draw, cylinder, shape border rotate=90, anchor = west, minimum height=12.5em, minimum width=3em] 		(0)		{$\text{DB}$};

\node[block, align=center]							(B0)[right of=0, yshift=.75em,  xshift=1em ]		{$M_2$};
\node[block, anchor= north, align=center]			(A0)[above of=B0, yshift=0.8em ]	{$M_1$};
\node[block, anchor = south, align=center,label=below:{\color{white}'\color{black}$\sigma_{i,1}$\color{white}'}]			(C0)[below of=B0, yshift=-0.8em]	{$M_3$};

\node[draw=none, fill=none]    		(XB1)[right of=B0, xshift=1em]	{\textbf{...}};
\node[draw=none, fill=none]    		(XA1)[right of=A0, xshift=1em]	{\textbf{...}};
\node[draw=none, fill=none]    		(XC1)[right of=C0, xshift=1em]	{\textbf{...}};

\node[block, align=center]							(B3)[right of=XB1,  xshift=.75em ]		{$M_2$};
\node[block, anchor= north, align=center]			(A3)[above of=B3, yshift=0.8em ]	{$M_1$};
\node[block, anchor = south, align=center,label=below:{\color{white}'\color{black}$\sigma_{i,r}$\color{white}'}]			(C3)[below of=B3, yshift=-0.8em]	{$M_3$};

\node[block, align=center]							(B4)[right of=B3,  xshift=0.5em ]		{$M_2$};
\node[block, anchor= north, align=center]			(A4)[above of=B4, yshift=0.8em ]	{$M_1$};
\node[block, anchor = south, align=center]			(C4)[below of=B4, yshift=-0.8em]	{$M_3$};

\node[block, align=center]							(B5)[right of=B4,  xshift=0.5em ]		{$M_2$};
\node[block, anchor= north, align=center]			(A5)[above of=B5, yshift=0.8em ]	{$M_1$};
\node[block, anchor = south, align=center]			(C5)[below of=B5, yshift=-0.8em]	{$M_3$};

\node[block, align=center]							(B6)[right of=B5,  xshift=0.5em ]		{$M_2$};	
\node[block, anchor= north, align=center]			(A6)[above of=B6, yshift=0.8em ]	{$M_1$};
\node[block, anchor = south, align=center,label=below:{$\sigma_{i,1}'$}]			(C6)[below of=B6, yshift=-0.8em]	{$M_3$};

\node[draw=none, fill=none]    		(XB7)[right of=B6, xshift=1em]	{\textbf{...}};
\node[draw=none, fill=none]    		(XA7)[right of=A6, xshift=1em]	{\textbf{...}};
\node[draw=none, fill=none]    		(XC7)[right of=C6, xshift=1em]	{\textbf{...}};

\node[block, align=center]							(B9)[right of=XB7,  xshift=.75em ]		{$M_2$};
\node[block, anchor= north, align=center]			(A9)[above of=B9, yshift=0.8em ]	{$M_1$};
\node[block, anchor = south, align=center,label=below:{$\sigma_{i,r}'$}]			(C9)[below of=B9, yshift=-0.8em]	{$M_3$};

\node[draw, cylinder, shape border rotate=90, anchor = east, minimum height=12.5em, minimum width=3em] 	(00)[right of=B9, xshift=1em, yshift=-.75em]				{$\text{DB}$};

\draw[decoration={brace, amplitude=1em}, decorate]  ([yshift=2em,xshift=-1em]A0.north east) -- ([yshift=2em, xshift=1em]A3.north west) node[above, pos=0.5, yshift=1em] {Unwrapping phase};

\draw[decoration={brace, amplitude=1em}, decorate]  ([yshift=2em,xshift=-1em]A3.north east) -- ([yshift=2em,xshift=1em]A6.north west) node[above, pos=0.5, yshift=1em] {Simple E/D phase};

\draw[decoration={brace, amplitude=1em}, decorate]  ([yshift=2em,xshift=-1em]A6.north east) -- ([yshift=2em,xshift=1em]A9.north west) node[above, pos=0.5, yshift=1em] {Wrapping phase};

\draw[decoration={brace, amplitude=0.5em}, decorate]([yshift=12em]0.east) -- ([yshift=6em]0.east) node {};
\draw[decoration={brace, amplitude=0.5em}, decorate]([yshift=4.5em]0.east) -- ([yshift=-1.5em]0.east) node {};
\draw[decoration={brace, amplitude=0.5em}, decorate]([yshift=-3em]0.east) -- ([yshift=-9em]0.east) node {};

\draw[decoration={brace, mirror, amplitude=0.5em}, decorate]([yshift=12em]00.west) -- ([yshift=6em]00.west) node {};
\draw[decoration={brace, mirror, amplitude=0.5em}, decorate]([yshift=4.5em]00.west) -- ([yshift=-1.5em]00.west) node {};
\draw[decoration={brace, mirror, amplitude=0.5em}, decorate]([yshift=-3em]00.west) -- ([yshift=-9em]00.west) node {};

\path[->, midway]
 ([yshift=8.9em,xshift=1em]0.east) edge node				{} (A0.west)
 ([yshift=1.5em,xshift=1em]0.east) edge node						{} (B0.west)
 ([yshift=-5.9em,xshift=1em]0.east) edge node				{} (C0.west)

 (A0.east) edge node[above,yshift=0.5em]	{$\sigma_{pub,1}$} 	(XA1)
 (A0.east) edge node						{} 					(XB1)
 (A0.east) edge node						{} 					(XC1)
 (B0.east) edge node						{} 					(XA1)
 (B0.east) edge node						{} 					(XB1)
 (B0.east) edge node						{} 					(XC1)
 (C0.east) edge node						{} 					(XA1)
 (C0.east) edge node						{} 					(XB1)
 (C0.east) edge node						{} 					(XC1)
 
 (XA1.east) edge node[above,yshift=.5em]	{$\sigma_{pub,r-1}$} 	(A3)
 (XA1.east) edge node						{} 					(B3)
 (XA1.east) edge node						{} 					(C3)
 (XB1.east) edge node						{} 					(A3)
 (XB1.east) edge node						{} 					(B3)
 (XB1.east) edge node						{} 					(C3)
 (XC1.east) edge node						{} 					(A3)
 (XC1.east) edge node						{} 					(B3)
 (XC1.east) edge node						{} 					(C3)

 (A3.east) edge node						{}				 	(B4)
 (B3.east) edge node						{} 					(C4)
 (C3.east) edge node						{} 					(A4)
 (A4.east) edge node						{}				 	(B5)
 (B4.east) edge node						{} 					(C5)
 (C4.east) edge node						{} 					(A5)
 (A5.east) edge node						{}				 	(B6)
 (B5.east) edge node						{} 					(C6)
 (C5.east) edge node						{} 					(A6)
 
 (A6.east) edge node[above,yshift=.5em]	{$\sigma_{pub,1}'$}		(XA7)
 (A6.east) edge node						{} 					(XB7)
 (A6.east) edge node						{} 					(XC7)
 (B6.east) edge node						{} 					(XA7)
 (B6.east) edge node						{} 					(XB7)
 (B6.east) edge node						{} 					(XC7)
 (C6.east) edge node						{} 					(XA7)
 (C6.east) edge node						{} 					(XB7)
 (C6.east) edge node						{} 					(XC7)

 (XA7.east) edge node[above,yshift=.5em]	{$\sigma_{pub,r-1}'$}	(A9)
 (XA7.east) edge node						{} 					(B9)
 (XA7.east) edge node						{} 					(C9)
 (XB7.east) edge node						{} 					(A9)
 (XB7.east) edge node						{} 					(B9)
 (XB7.east) edge node						{} 					(C9)
 (XC7.east) edge node						{} 					(A9)
 (XC7.east) edge node						{} 					(B9)
 (XC7.east) edge node						{} 					(C9)
 
 (A9.east) edge node						{} 					([yshift=8.9em,xshift=-1em]00.west)
 (B9.east) edge node						{} 					([yshift=1.5em,xshift=-1em]00.west)
 (C9.east) edge node						{} 					([yshift=-5.9em,xshift=-1em]00.west);

\end{tikzpicture}
\centering
\caption{Parallel Mix-ORAM with 3 mixes.\newline Rebuild method (all phases) and Layered method (only the Wrapping phase).}\label{fig:Par}
\end{figure*}
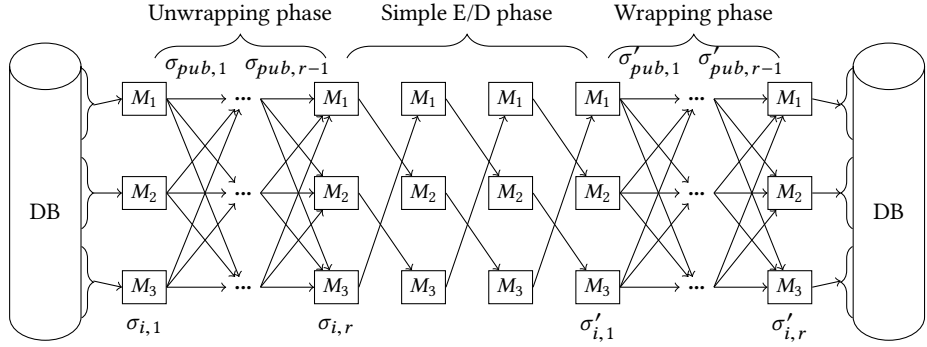

We now consider the shuffling methods with the mix-net in a stratified configuration, where all the mixes perform the same operations in parallel and forward the output to each other as shown in Figure~\ref{fig:Par}.

The mixes have each been allocated a chunk of the database ($M_{idx}$ having $[idx\cdot n/m : (idx+1)\cdot n/m]$) and use the public permutation seeds $\sigma_{pub}$ to compute which records to send to each mix.

\subsubsection{Parallel Layered scheme.}
In this design, depicted as the Wrapping phase of Figure~\ref{fig:Par}, the records are still appended with a label and an IV token, encrypted and permuted as in subsection~\ref{section:CL}, however now chunks of the database are assigned and processed by each mix. Before the eviction, the database is permuted with the old seeds $\sigma_i$ and encrypted with the old keys $k_i$. Afterwards, the records are encrypted with both $k_i$ and $k_i'$, permuted with both $\sigma_i$ and $\sigma_i'$, and the new indices are saved on the client. 
As no permutation layer is ever removed, the record indistinguishability assumption holds, the eviction then consists of $r= m/2 \ \log(n/s)$ rounds. \\

\noindent\textit{Mix Instructions.}
The client needs to send to each mix the session keys to access the database $db$, the private and public elements used to compute the encryption keys, the permutation seeds and the record allocation $\alpha_i$ and $\beta_i$, the security parameter $\kappa$, the number of records and rounds $n$ and $r$, and the ordered $list=(ports,\ ips)$ of the mixes participating in the eviction. The client thus send :
$$C \rightarrow M_i\ :\ db,\ \alpha_i,\ \beta_i,\ \kappa,\ n,\ r,\ list$$

\noindent We generate the permutation seeds and encryption keys as before and furthermore refresh them at each round by blinding the group elements. Let $h_b : \mathcal{G}^* \rightarrow \mathbb{Z}_{q}^*$ be the hash function used for computing blinding factors, we can then compute recursively the $\alpha$ and $\beta$ for the round $j+1$ as follows:
\begin{align*}
\alpha_{i,0} &= g^{z_i}, &ss_{i,0 }&= y_{i,0}^{z_i}, &k_{i,0},\ \sigma_{i,0}&=hkdf(ss_{i,0},\ \kappa)\\
\beta_{i, 0} &= g^{\Pi_{j\neq i}m_j}, &sk_0 &= \beta_{i,0}^{m_i}, &\sigma_{pub, 0}&=hkdf( sk_0,\ \kappa)
\end{align*}
\vspace{-.5cm}
\begin{align*}
b_{i,j+1}&=h_b(\alpha_{i,j}, ss_{i,j}), & \alpha_{i,j+1} &= g^{z_i\Pi_{k\leq j}b_{i,k}}\\
b_{pub,j+1}&=h_b(sk_{j}), &\beta_{i, j+1} &= g^{\Pi_{k\leq j}b_{pub,k}\Pi_{l\neq i}m_l}\\
\end{align*}

\noindent\textit{Mix operations.} In this scheme, the mix $M_i$ receives a list of encrypted record from all mixes or from the database at the first round. It first merges the records and sorts them to the order given by the previous public record allocation. It then encrypts each record and permutes them with the private encryption key and private permutation seed. It finally blinds the public permutation seeds, computes the new record allocation and sends the records to the mixes accordingly, or to the database at the last round.\\

\noindent\textit{Client operations.} The client can find the record index locally as it was stored before the eviction. To decrypt a record, the client uses a  similar algorithm to the one used in the cascade configuration. The Parallel Trial and Error algorithm, as described in Algorithm~\ref{alg:ptrial}, now needs to decrypt the records for each eviction and each rounds, and determines for each round which mix processed the desired record. The access method is the same as in the Cascade. \\

\begin{algorithm}
\SetKwComment{Comment}{//}{}
\DontPrintSemicolon
\KwIn{Record and index $rec,\ index$;  \\ \qquad Shared and private encryption keys $k_{mix,eviction,round},\ prv$;\\ \qquad Permutation seeds $\sigma$;\\ \qquad Number of rounds $r$; \\ \qquad List of mixes $list= (ips,\ ports)$;}
$j,e=0$;\\
$r\gets \text{decrypt}(prv,rec)$;\\
\While{$rec.data.label != i$}
{
\If {$e!=0$}
{
$rec \gets \text{encrypt}(prv,\ rec)$;\\
}
\ForAll{$j \in \llbracket 1 :r \rrbracket $}
{
	$m\gets \text{retrieve\_mix}(\sigma,\ e,\ j,\ index,\ list)$;\\
	$rec \gets \text{decrypt}(k_{m,\ e,\ j}, rec)$;\\
	$j\gets j-1$;\\
}
	$rec \gets \text{decrypt}(prv,\ rec)$;\\
    $e\gets e-1$;\\
}
\KwOut{rec}
\caption{Parallel Layered Trial Error Algorithm}
\label{alg:ptrial}
\end{algorithm}

\noindent\textit{Costs.} The mix communication cost is $ (r+2) \cdot n \cdot b$, the mix permutation cost is $m \log(n/s) \cdot C_{\Pi}(n/m)$ with $C_{\Pi}(n/m)$ the cost of permuting $n/m$ elements and the encryption cost $m/2 \log(n)\cdot  C_{cbc}$ with $C_{cbc}$ the cost of encrypting one data block. The client Lookup cost is of the order $O(1)$ as the indices, i.e. $n\log n$ bit, are stored locally. The client decryption cost will be talked in Section~\ref{Evaluation}, and the group elements stored on the client represents $2\kappa m$ bits.

\subsubsection{Parallel Rebuild method.}
This design, depicted in Figure~\ref{fig:Par}, is composed of three phases as in the Cascade configuration. However, we now assign a chunk of the database to each mix which processes during a specified of rounds $r = 2n/k \log(n)$ during each permutation phase and encrypt some of the records in parallel. Before the eviction, the records are permuted and encrypted by the mixes $M_i$ with the permutation seeds $\sigma_{pub}$ and $\sigma_i$ and the encryption keys $k_i$. Afterwards, the database is encrypted with the keys $k_i'$ and permuted with the seeds $\sigma_{pub}'$ and $\sigma_i'$.\\

\noindent\textit{Mix Instructions.}
The client sends the same information as in the Parallel Layered design but with twice the number of group elements: 
$$C \rightarrow M_i\ :\ db,\  \alpha_i,\ \alpha_i',\ \beta,\ \beta',\ \kappa,\ n,\ r,\ list$$

To derive the permutation seeds $\sigma$ and encryption keys $k$, we make use of the random private elements $z,i$ and $m_i$, the public and private keys $y$ and $x$ as in the Layered method. We furthermore derive the $\alpha$ $r$ more times for the simple E/D phase. 
\begin{align*}
b_{i,j+1}&=h_b(\alpha_{i,j}, ss_{i,j}), & \alpha_{i,j+1} &= g^{z_i\Pi_{k\leq j}b_{i,k}}\\
b_{pub,j+1}&=h_b(y_c, sk_{j}), &\beta_{i, j+1} &= g^{\Pi_{k\leq j}b_{pub,k}\Pi_{l\neq i}m_l}\\
\end{align*}

\vspace{-1em}

\noindent\textit{Mix Operations.} During the first $r$ rounds, the records are first sorted according to the previous public record allocation, then unwrapped (permuted and decrypted with the old keys and seeds) and sent to the mix-net according to the public record allocation generated from the blinded public seeds. Then the groups of $n/m$ records are encrypted and decrypted in $m$ parallel cascades. Finally, the records are similarly sorted, wrapped (encrypted and permuted with the new keys and seeds) and sent to the mix-net during the last $r$ rounds.\\

\noindent\textit{Client Operations.} To find a record position, the client uses a similar algorithm as the one used in the Cascade Rebuild scheme. The Parallel Index Lookup Algorithm, as described in Algorithm~\ref{alg:PIL}, however needs to determine at which round where the record was sent and processed and compute the associated keys and seeds. These intermediary results, the list of indices, keys and seeds, can be stored to facilitate the decryption of the record ; the method being similar to the one used in the Cascade configuration.\\

\begin{algorithm}
\SetKwComment{Comment}{//}{}
\DontPrintSemicolon
\KwIn{Private and public seeds $\sigma_{i,round},\ \sigma_{round}$;\\ \qquad Number of records, mixes and rounds $n,\ m,\ r$;\\ \qquad Record index $index$;}
$mixes \gets \{,\}$;\\
$indices \gets \{,\}$;\\
\ForAll {$i \in \llbracket 1, r \rrbracket$}
{
	$mix \gets \floor{index/m}$;\\
	$mixes \gets mixes \cup \{mix \}$;\\
	$shuffle \gets {\Pi}_{\sigma_{mix,i}}(i\cdot n / m,\ )i+1)\cdot n / m)$;\\
	$index \gets i\cdot n/m + shuffle.index(index)$;\\
	$indices \gets \cup \{index\}$;\\
	$shuffle \gets {\Pi}_{\sigma_{i}(1,n)}$;\\
	$index \gets shuffle.index(index)$;\\
}
\KwOut{$mixes,\ indices$}
\caption{Parallel Index Lookup}
\label{alg:PIL}
\end{algorithm}

\noindent\textit{Costs.} The mix communication cost is $ (2m+r+2) \cdot n \cdot b$, the mix permutation cost is $8m \log n \cdot C_{\Pi}(n/m)$ with $C_{\Pi}(n/m)$ the cost of permuting $n/m$ elements and the encryption cost $n (4 \log n +2) \cdot C_{ctr}$ with $C_{ctr}$ the cost of encrypting one data block. The client Lookup cost is of the order $m (C_{\Pi}(n)+C_{\Pi}(n/m))$. The client decryption cost is $(r+m) C_{ctr}$, and the group elements stored on the client represents $2\kappa (m+1)$ bits.

\section{Security Argument}\label{Security}

We first remark that all of the eviction meta data is independent of data content, as it is entirely determined by the sole parameter $n$. The mix instructions are never shared between parties, the keys and seeds thus remain secret and are refreshed at every round.\\

\noindent\textbf{Cascade mix-net.}
In this architecture, the whole database passes by every mix including the honest one where it is locally permuted and re-encrypted with the private shared keys. As a polynomial adversary cannot break the PRF, the database order is kept confidential.

For the \emph{Rebuild method}, the simple Encryption/Decryption phase ensures that the records are always encrypted as the adversary is not able to break the double AES encryption.\\

\noindent\textbf{Parallel mix-net.}
In this architecture, chunks of the database are exchanged between mixes during $r$ rounds. The adversary can benefit of the fact that some records may never go to the honest mixes but this happens with negligible probability of $p=(e^{-r/m} << 1$ with our parameters.

For the \emph{Rebuild method}, we derived the number of rounds needed to secure our design from the method used in Goodrich 2012~\cite{goodrich2012anonymous} to quantify the information leakage (see Appendix~\ref{proof:pmn}) and found that this number of rounds is sufficient to bound the expected sum of square error between the card assignment probabilities and the uniform distribution by at most $1/n^2$. The simple E/D phase similarly to the Cascade configuration prevents any mix or the adversary from the decrypting the records completely when they are not permuted and thus the leakage of which records were accessed.

For the \emph{Layered method}, we proposed to use the previous randomness to reduce the number of rounds needed to be close to the uniform distribution. We can also reuse Goodrich's proof by changing the probabilities such that $w_i(t)$ being now the probability the $i^{th}$ record at the $t^{th}$ round was in the cache at first and $\Phi(t)=\Sigma w_i(t) - s/n$, we obtain $r>m\log(n/s)$ see Appendix~\ref{proof:OM} and Conjecture~\ref{sec:kOM}).

\section{Evaluation}\label{Evaluation}
\textbf{Layered method.} We look here at the average number of encryption layers $e$ a record has before being decrypted. Making the assumption that the record access distribution is uniform, we can represent the problem of accessing all records at least once as a coupon collector problem. In that case, the average number of evictions before all records have been fetched once is $E[e_{all}]\leq(n/s)\cdot H_n$ with $H_n$ the $n^{th}$ harmonic number. The expected number of encryption layers per record before decryption is however $E[r]\leq{r/s} \cdot \left ( \frac{n+1}{2}\cdot(H_n-1/2)+1/2 \right )$. For $n=10^6$ and $s=\sqrt{n}$, we have $E[e]\approx 15 \cdot 10^3$ and $E[r]\approx 7\cdot 10^3 \cdot r$.
\begin{proof}
Lets $\tau_n$ be the random number of coupons collected when the first set contains every $n$ types. We have, $E[\tau_n]=n\sum_{i=1}^n \frac{1}{i} = n \cdot H_n$.
Since we fetch $s$ unique records per eviction (we cannot fetch a record already in the stash), the previous result is an upper bound of the number of requests needed and so the expected number of eviction is $E[e_{all}]\leq n/s\ H_n$.

We now want to find the average number of encryption layers per record before decryption, this is equivalent to finding the average number of evictions before a record is deciphered. 
Hence we have, $E[e]\leq r/s \cdot \sum_{i=1}^n E[\tau_i] = r/s\ \sum_{i=1}^n \left (\frac{(n+1-i)(n+i)}{2}\cdot \frac{1}{i}\right )$ from which can be calculated the result presented earlier. 
\end{proof}

To reduce these numbers, we can modify the access method as follows. When the client requests a record from the database, $d$ other records are chosen uniformly at random from the set of unaccessed records. These records are then fetched, their encryption is refreshed as written previously and the client overwrites with these records their older version on the database. Doing so, with $d$ high enough, yields a better approximation of the uniform distribution assumption and we would obtain  $E[e_{all}]\leq n/(sd) \cdot H_n $ and $E[e] \leq {r/(sd)} \cdot \left [ \frac{n+1}{2}\cdot(H_n-1/2)+1/2 \right ] \ H_n$.
With $d=\sqrt n$, we now have $E[e_{all}]\leq 15 r$ and $E[e]\leq 7r$.

Another method to reduce the decryption cost would be to reinitialize the database periodically, for instance every $e$ eviction. Doing so, the client would only need to decipher each record a maximum of $e\cdot m$ times for the Cascade architecture and $e \cdot r$ times for the Parallel architecture during the reinitialization process and in the decryption method.

\section{Comparison}\label{Comparison}
We can find in Table~\ref{tab:CComp} and Table~\ref{tab:PComp} the cost comparisons of the different Mix-ORAM designs. We did not include the public permutation costs in the Parallel cases as they can be done offline, or during the records' exchange at each round, since the permutation is done on the range of indices and not on the data.
We can see that the Layered method is more efficient than the Rebuild one in theory, however we have to take into account in practice the added cost due to the fact that the whole database may not fit in the cache of the mixes (the time needed to fetch the records from the main memory). Moreover, the client incurs higher costs, both in term of memory and computation, with the Layered method. 
Comparison of our schemes.
The mix cost in the Cascade architecture are quadratic in the number of mixes $m$ while they could be considered independent of $m$ in the Parallel case. Hence, the Parallel architecture, even if it has a higher number of rounds, can still be faster than the Cascade depending on the siwe of the network and of the cache for the Layered method.\\

\begin{table*}
\centering
\begin{tabular}{l *2c}
\toprule
    					& Cascade - Layered	 			& Cascade - Rebuild						\\
\midrule
Mix memory & $n$ & $n$ \\
Mix Encryption cost & $m \cdot  n$ & $4m  \cdot  n$  \\
Mix Permutation cost & $m n \cdot C_{\Pi}(n)$ & $2 m n C_{\Pi}(n)$ \\
Mix Communication cost & $(m+1) \cdot C_{com}(n)$ & $3m \cdot C_{com}(n)$\\
Client Lookup overhead & $O(1)$ & $m\cdot C_{\Pi}(n)$ \\
Client Decryption overhead & $\sim \frac{nm}{2s} H_n$ & $2m$ \\
Client Storage overhead & $n\log(n)+ 2 \kappa m$ & $2 \kappa m$ \\
\bottomrule
\end{tabular}
\centering
\caption{Cost comparison of the designs with $C_{E}$ the cost of 1 encryption, $C_{\Pi}(x)$ the permutation cost and $C_{com}(x)$ the communication cost of $x$ records in the scheme.}
\label{tab:CComp}
\end{table*}

\begin{table*}
\centering
\begin{tabular}{l *4c}
\toprule
& Parallel - Layered 						& Parallel - Rebuild\\
\midrule
\#Rounds ($r$)  & $\frac{m}{2} \log\left( \frac{n}{s}\right)$ &  $2m\log(n) $ \\
Mix memory & $n/m$ & $n/m$ \\
Mix Encryption cost  & $\frac{n}{2} \log\left (\frac{n}{s}\right)$ & $n(4\log(n) + 2) $ \\
Mix Permutation cost & $m \log\left (\frac{n}{s}\right)\cdot C_{\Pi}\left(\frac{n}{m}\right)$ & $8m\log(n) \cdot C_{\Pi}\left ( \frac{n}{m} \right )$ \\
Mix Communication cost & $m(r+2) \cdot C_{com}\left(\frac{n}{m}\right)$ & $m(2r+m+2) \cdot C_{com}\left(\frac{n}{m}\right)$\\
Client Lookup overhead & $O(1)$ & $m \cdot [C_{\Pi}\left ( \frac{n}{m}\right )h+ 2C_{\Pi}(n)]$\\
Client Decryption overhead &  $\sim \frac{nr}{2s} H_n$ & $m+r$\\
Client Storage overhead &$n\log(n)+ 2 \kappa  (m+1)$ & $2 \kappa  (m+1)$ \\
\bottomrule
\end{tabular}
\centering
\caption{Cost comparison of the designs with $C_{E}$ the cost of 1 encryption, $C_{\Pi}(x)$ the permutation cost and $C_{com}(x)$ the communication cost of $x$ records in the scheme.}
\label{tab:PComp}
\end{table*}

Comparing the computation and communication costs of our designs to existing eviction schemes would be interesting but delicate as we take into account the fact that the mixes may have faster processor or larger RAM and that the bandwidth in the mix-net may be higher than the one between the client and the ORAM process thus speeding the eviction.
The total communication or computation cost of each of our design is indeed higher than regular evictions' such as Melbourne's~\cite{ohrimenko2014melbourne}. However, in our cases, the client only needs to preprocess the database once, and with the periodic reinitialization of the database in the Layered approach, and has manageable additional costs for the lookup and description of a record. These costs do not compare with the overhead of the periodic eviction incurred by the client in the non delegated schemes eviction. 

\section{Acknowledgement}
Danezis was supported by H2020  PANORAMIX Grant (ref. 653497) and EPSRC Grant EP/M013286/1; and Toledo by Microsoft Research.

\section{Conclusion}\label{Conclusion}
We presented in this paper a novel ORAM eviction system where the randomization, more specifically the oblivious shuffle, is delegated to a semi-trusted mix-net. Doing so, the client is alleviated from the main overhead of the ORAM technology at the cost of reasonable additional costs for the record lookup and decryption. Very thin clients can thus accede to the ORAM technology as only a few group elements are needed for fetching any records. The database is moreover accessible and can be made available during the eviction of the records, and this independent of the structure of the underlying ORAM server, making the ORAM technology more portable.

\newpage
\bibliographystyle{ACM-Reference-Format}
\bibliography{mix}

\newpage
\newpage
\section{Appendix}
\subsection{Proof $k$-RTS}\label{proof:kRTS}
\begin{proof}
To prove the upper bound, we use Diaconis et al. method ~\cite{aldous1986shuffling} which consists in marking cards depending on whether they have already been picked or not. Let's define $\tau$ the stopping time, i.e. the time when every card has been marked and $\tau_i$ the number of transpositions before $i$ cards have been marked. The $\tau_i$ are independent geometric variables with probability of success $p_t$ as implied by the game rules.
We thus have,
\begin{align*}
 p_t &= \sum_{i=1}^{min(k,n-t)} {k \choose i} \cdot {t+1 \choose i} \cdot {n-t \choose i}\cdot{n \choose i}^{-2}&\\
 &= \frac{1}{n^2} \cdot \left ( k \cdot (t+1)\cdot(n-t) + \alpha_{n,t,k}\right )
\end{align*}
With $\alpha_{n,t,k} = \mathcal{O}\left(n^{-k}\right) $ positive.\\\

\noindent We can thus rewrite $\tau$'s expectation as following.
\begin{align*}
 E(\tau) &=  E \left ( \sum_{i=0}^{n-1} \tau_i \right ) = \sum_{t=0}^{n-1} \frac{1}{p_{t}}  < \sum_{t=0}^{n-1} \frac{n^2}{k \cdot (t+1)\cdot(n-t)}&\\  
 &< \frac{2}{k} \cdot \frac{n^2}{n+1} \cdot \left( \ln(n) + \gamma +\mathcal{O}(\frac{1}{n}) \right),\ \gamma = \lim_{n \to \infty} H_n - \ln(n)& \\
 \end{align*}
 \end{proof}
 
\subsection{Proof of Oblivious Merge}\label{proof:OM}
\begin{proof}
We want to find the mixing time $\tau(\epsilon)$ of our oblivious merge of two sets of indistinguishable elements. To do so, we use the bound of the mixing time of an irreducible ergodic Markov Chain, where $p = \frac{1}{|V|}$, with the volume $V={n \choose s}$, and $1-\lambda^*$ is the spectral gap, we thus have,
$$\frac{\lambda^*}{1-\lambda^*} \cdot \log\left(\frac{1}{2 \epsilon} \right)\leq \tau(\epsilon) \leq \frac{1}{1-\lambda^*}\cdot \log \left( \frac{1}{2 \epsilon \cdot \sqrt{p}}\right) $$

We now represent the arranging of merge of the 2 distinct sets by the graph $\mathcal{G}$, a $k$-regular graph with $v$ vertices corresponding to the different orderings and the undirected edges to transpositions of two elements.
By definition, the eigenvalues of the transition matrix of the $\mathcal{G}$ are $k={\lambda'}_0 > {\lambda'}_1 \geq  ... \geq {\lambda'}_{n-1}$, and we have,
$$\text{diam}\left( \mathcal{G}\right) \leq \frac{log(v-1)}{log(\frac{k}{{\lambda'}^*})}+1 \text{ with } {\lambda'}^* = max_{i\neq0}({\lambda'}_i)= k \cdot \lambda^*$$
From which we can deduce that $ {\lambda}^* \geq \left ({n \choose s}-1 \right )^{\frac{1}{1-s}} \geq \left (\frac{n\cdot e}{s} \right )^{\frac{s}{1-s}} $ since $diam\left( \mathcal{G} \right)=s$ the diameter of the graph, $v= {n \choose s}$ the number of vertices and $k=s\cdot(n-s)$.

To find an upper-bound of $\lambda^*$, we will now look at spectral gap bounding.
Let's $\mathcal{G}_{0,1}=\{0,1\}^n$ be the group of elements with the XOR operation and $\mathcal{S}=\{x \in \mathcal{G},\ weight(x)=s\}$ the symmetric subset of $\mathcal{G}$ of n-binary array with $s$ 1s and $n-s$ 0s.
We call $Cay_{n,s}=Graph\left(  \mathcal{G}_{0,1}, \mathcal{S} \right) $ the Cayley graph generated from these structures.

\begin{lemma}
Let $\mathcal{G}$ be a finite Abelian group, $\chi\ :\  \mathcal{G} \rightarrow \mathbb{C}$ be a character of $\mathcal{G}$, $\mathcal{S} \subseteq \mathcal{G}$ be a symmetric set.
Let $M$ be the normalized adjacency matrix of the Cayley graph $G = Cay(\mathcal{G},\mathcal{S})$.
Consider the vector $x \in \mathbb{C}^\mathcal{G}$ such that $x_a = \chi(a)$. Then x is an eigenvector of $G$, with eigenvalue $$ \frac{1}{\mathcal{S}} \sum_{s\in \mathcal{S}} \chi\left(s\right)$$
\end{lemma}

\begin{theorem}
The Cayley graph $Cay_{n,s}$ has for eigenvalues $\mu_0 = 1 > \mu_1 \geq ... \geq \mu_{n-1}$ with, 
$$\mu_r = \frac{1}{\left | \mathcal{S} \right |} \sum_{i=1}^{min(r, n-r)} \left ( -1 \right )^i {r \choose i}{n-r \choose s-i}\\$$
\end{theorem}

\begin{subproof}
$\forall r \in \{0,1\}^n$, with $\chi_r(x)=\left ( -1 \right )^{\sum r_i \cdot x_i}$, we have,
\begin{align*}
\mu_r &= \frac{1}{\mathcal{S}} \sum_{s\in \mathcal{S}} \chi\left(s\right) = \frac{1}{\left | \mathcal{S} \right |} \sum_{s\in \mathcal{S}} \left ( -1 \right )^{\sum r_i \cdot s_i}\\
& = \frac{1}{\left | \mathcal{S} \right |} \sum_{i=1}^{min(r, s)} \left ( -1 \right )^i {r \choose i}{n-r \choose s-i} \\
\end{align*}
\end{subproof}

Remark. We recognize here the Vandermonde identity with alternating numbers. We argue that the eigenvalues of the Cayley graph $Cay_{n,s}$ are all positive as the smallest eigenvalue is null.
For $r=n-r$, the expression simplifies to $\mu_r = {r \choose \frac{n}{2}}$ if $n$ even, 0 otherwise.
For $r=1$, the expression simplifies to $\mu_1 = 1 - 2\cdot \frac{s}{n}$, the spectral gap of $Cay_{n,s}$ is thus equal to $2\cdot \frac{s}{n}$.\\\

We notice that the first graph $\mathcal{G}$ actually is a sub-graph of $Cay_{n,s}$ and as such the adjacent matrix of the first graph is included in the second's.
For $s>1$, $Cay_{n,s}$ is divided in two sub-graphs representing the cosets of $\{0,1\}^n$ as $\mathcal{S}$ is not a generating group of $\mathcal{G}_{0,1}$, $\mathcal{G}$ is only contained in one of the sub-graphs.
We use the Cauchy's Interlace Theorem to bound the eigenvalues of $\mathcal{G}$ with the ones of $Cay_{n,s}$,.

\begin{theorem}
Let $M$ be a Hermitian $n \times n$ matrix with eigenvalues ${\mu'}_0\geq ... \geq {\mu'}_{n-1}$ and $N$ a $m \times m$ sub-matrix of $M$ with eigenvalues ${\lambda'}_0\geq ... \geq {\lambda'}_{m-1}$ , we have
$$ {\mu'}_i \geq {\lambda'}_i \geq {\mu'}_{n-m+i+1} $$
\end{theorem}

We are here only interested in an upper-bound of $\lambda*$, as we have $\mu_{2^n+2-{n \choose s}}\leq \lambda_1\leq 1-2\frac{s}{n}$ and $0 \leq \lambda_n \leq \mu_2$, $\lambda* \leq 1-2\frac{s}{n}$. We thus have $\frac{1}{1-\lambda*}\leq\frac{n}{2\cdot s}$ and $\log {n \choose s} \approx s(\log(n/s-0.5)+1) -1/2\log(2\pi s)$ when $n \gg s$ from which we derive the final result.
\end{proof}

\subsection{Proof of Parallel mix-net}\label{proof:pmn}
\begin{proof}
This proof is derived from Goodrich et al~\cite{goodrich2012privacy} who bounded the closeness of a shuffle to the uniform distribution using a compromised parallel mix-net. 

Let $w_i(t)$ the probability the $i^{th}$ record at the $t^{th}$ round was the first record at start, the sum of square metric $\Phi(t)=\Sigma_{i=1}^n (w_i(t)-1/n)^2$, $n$ the number of cards, $m$ the number of mixes out of which $m_a$ are corrupted and $k=n/m$.

We have by recurrence that the potential $\Delta\Phi^*$ changes when a group of K card is shuffled during a round as following : $\Delta\Phi^*=\Sigma_{1\leq i\leq n}(w_i-w_j)^2$.
Thereby,

\begin{align*}
E[\Delta\Phi] & = \frac{m}{n} \sum_{1\leq i\leq n} \Pr((i,j)\ in\ the\ same\ honest\ mix) (w_i-w_j)^2\\
&=\frac{k-1}{k(n-1)}\cdot \frac{m-m_a}{m} \sum_{i<j}(w_i-w_j)^2 \\\
E[\frac{\Delta\Phi}{\Phi}]&=\frac{(m-m_a)(k-1)}{2n(n-1)}\frac{\sum_{i,j}((w_i-1/n)-(w_j-1/n))^2}{\sum_k (w_k-1/n)^2}\\
&=\frac{(m-m_a)(k-1)}{n-1} \text{ since $\sum_k w_k -1/n=0$}
\end{align*}

We thus find,
\begin{align*}
E[\Phi(t+1)] &= (1-\frac{(m-m_a)(k-1)}{n-1}) E[\Phi(t)]\\
E[\Phi(t)]&= (1-\frac{(m-m_a)(k-1)}{n-1})^t
\end{align*} 

We want to find the conditions on $c$ such that the corrupted parallel mix-net can mix in $t=bc\log(n)$  such that $E[\Phi(t)]< n^{-b}$.

\begin{align*}
E[\Phi(t)] = (1-\frac{(m-m_a)(k-1)}{n-1})^t &< n^{-b}\\
c \cdot\log(1+ \frac{1}{\frac{n-1}{(m-m_a)(k-1)}- 1}) &> 1\\
\end{align*}

Using Taylor series, assuming that $n-1\gg(m-m_a)(k-1)$,  we finally get
\begin{align*}
&c \cdot (\frac{1}{\frac{n-1}{(m-m_a)(k-1)}- 1} +o(n/k)) > 1 \\
&c > \frac{n-1}{(m-m_a)(k-1)}- 1 \simeq \frac{m}{m-m_a} -o(1)\\
\end{align*}
Thus, when shuffling $n$ cards with n a parallel mix-net composed of $m$ mixes out of which $m_a$ were compromised, we need $t>b\cdot \frac{m}{m-m_a} \log(n) $ rounds before the expected sum of squares error $E[\Phi(t)]$ between the card assignment probabilities and the uniform distribution is at most $1/n^b$ for any fixed $b>1$.
\end{proof}

\end{document}